\documentclass[aps,prb,twocolumn,groupedaddress,showpacs,amsmath]{revtex4}

\usepackage{graphicx}
\usepackage{latexsym}

\usepackage{epsfig}
\usepackage{amssymb}

\newcommand{\kets}[1]{\ensuremath{|#1\rangle}}
\newcommand{\bras}[1]{\ensuremath{\langle#1|}}

\newcommand{\up}{\ensuremath{\uparrow}}

\begin{document}

\preprint{}

\title{Exact ground states of quantum spin-2 models on the hexagonal lattice}

\author{Marc   Andr{\'e}   Ahrens}      \email[]{maa@thp.uni-koeln.de}
\author{Andreas    Schadschneider}      \email[]{as@thp.uni-koeln.de}
\author{Johannes     Zittartz}          \email[]{zitt@thp.uni-koeln.de}
\affiliation{Institute  for  Theoretical  Physics,  University  of
Cologne, 50937          K{\"o}ln,          Germany}
\date{\today}

\begin{abstract}
We construct {\em exact} non-trivial ground states of spin-2
quantum antiferromagnets on the hexagonal lattice. Using the
optimum ground state approach we determine the ground state in
different subspaces of a general spin-2 Hamiltonian consistent
with some realistic symmetries. These states, which are not of
simple product form, depend on two free parameters and can be
shown to be only weakly degenerate. We find ground states with
different types of magnetic order, i.e.\ a weak antiferromagnet
with finite sublattice magnetization and a weak ferromagnet with
ferrimagnetic order. For the latter it is argued that a quantum
phase transition occurs within the solvable subspace.

\end{abstract}


\maketitle


\section{Introduction}

Usually one faces the problem of finding the ground state and
excitations of a given (many-body) Hamiltonian that is supposed to
describe the physical problem of interest. The solution then
allows, at least in principle, to determine other quantities,
e.g.\ correlation functions. Usually an exact solution is not
possible, especially in dimensions $d\geq 2$, and one has to rely
on approximations (e.g.\ perturbation theory) or numerical
methods.

Another approach does just the opposite \cite{arovas}. It starts
with a given ground state $\kets{\psi_0}$ and tries to find the
corresponding Hamiltonian. This can be achieved by finding
positive-semidefinite operators that annihilate $\kets{\psi_0}$.
Then any Hamiltonian which is a sum of these operators has
$\kets{\psi_0}$ as a ground state.

The {\em optimum ground state (OGS)} approach developed in
[\onlinecite{KSZ93,BS95,NZ96}] allows for a {\em systematic}
construction of such ground states. For a given class of
Hamiltonians one searches for subspaces of the space of
interaction parameters where the determination of the ground state
can be reduced to a {\em local} problem. OGS are characterized by
the fact that they are simultaneously ground states of all {\em
local} interactions contained in the Hamiltonian. It should be
emphasized that despite this reduction to a local problem the
properties of the ground state still depend on the structure of
the lattice.

Generic examples are simple tensor product states like the fully
polarized ferromagnet $\kets{\up\up\cdots\up}$. However, the OGS
approach is more flexible and also allows the construction of
antiferromagnetic states that have a more complicated structure.
This is possible if the ground state of the local Hamiltonian is
degenerate.  For one-dimensional quantum spin chains the generic
realization of OGS are matrix-product ground states
\cite{KSZ91,KSZ92,FNW92}. These have been studied systematically
for spin-1 [\onlinecite{KSZ93}], spin-3/2 [\onlinecite{NZ96}] and
spin-2 [\onlinecite{ASZ02}] quantum spin chains revealing the
existence of various phases, e.g.\ different types of Haldane
phases or magnetically ordered states. Here all ground state
properties, especially correlation functions, can be evaluated
exactly in closed form. It should be mentioned that matrix-product
states also play an important role for the DMRG-technique
\cite{W92,DMRG,Scholl05} which can be interpreted as variational
method based on states of matrix-product form
\cite{OR95,OR97,DMNS98}.

However, the true power of the OGS approach is revealed by the
application to spin models in higher dimensions $d \geq 2$.
Technically this generalization is achieved by the \emph{vertex
state model} approach \cite{NKZ97}. Here to each vertex of a
vertex model a quantum state is assigned in addition to a
Boltzmann weight. This allows for a very efficient bookkeeping and
provides an elegant way of specifying the ground state. Ground
state properties like correlation functions are then determined by
the partition function of a {\em classical} vertex model
\cite{Bax82} on the {\em same} lattice.

This approach has already been applied successfully to spin-3/2
systems on the hexagonal lattice \cite{NKZ97} and spin-2 systems
on the square lattice \cite{NKZ00}. These models can  be
considered as anisotropic generalizations of the
valence-bond-solid (VBS) model of Affleck et al.\ \cite{AKLT88}
since for a generic VBS state the spin $S$ and the coordination
number $z$ of the lattice are related by $S=z/2$.

An important point is the introduction of anisotropic
interactions. This leads to whole families of solvable models and
allows to study the parameter-dependence of the ground state
properties, in contrast to the isotropic VBS-type models which are
usually solvable only at isolated points. Indeed in both cases
\cite{NKZ97,NKZ00} a quantum phase transition from a disordered to
a N\'eel-ordered phase was found. As mentioned above the
properties of these quantum-critical points are determined by the
critical behaviour of a {\em classical} vertex model on the {\em
same} lattice. A similar situation has recently been studied by
Ardonne et al.\cite{Ardonne04}. They considered the quantum dimer
model introduced by Rokhsar and Kivelson\cite{RK88}. For this
model the ground state can be determined exactly. It turns out to
be an {\em optimum} ground state of a very simple form, namely an
equal amplitude superposition of states connected by the dynamics.
On a square lattice the Rokhsar-Kivelson model is critical. This
allows also to study the influence of small perturbations on the
critical behaviour \cite{FHMOS04,Ardonne04}. For the
Rokhsar-Kivelson model it turns out that perturbations can be
relevant or irrelevant, depending on the lattice structure.

The paper is organized as follows. In Sec.~\ref{sec:method} we
give a rather general introduction to the OGS approach. In
Sec.~\ref{sec:model} the OGS for the spin-2 case on a hexagonal
lattice are determined. Here also the properties of these states
are discussed qualitatively. Sec.~\ref{sec:concl} contains
concluding remarks and an outlook on future work.


\section{Optimum ground states}
\label{sec:method}

In the following we describe the basic idea of the OGS approach
which will then be applied to spin-2 models on the hexagonal
lattice.

As mentioned before the OGS approach starts from a general class
of Hamiltonians determined by the physical problem and its
symmetries. First the most general Hamiltonian ${\cal H}$ with
nearest neighbor interactions consistent with the following
symmetries is constructed:
\begin{enumerate}
\item  homogeneity in real space, \item  parity invariance (i.e.
$h_{ij}=h_{ji}$), \item  rotational invariance in the $xy$-plane
of spin space, \item  spin-flip invariance.
\end{enumerate}
All of these requirements are quite natural for a two-dimensional
lattice. The Hamiltonian then has the general structure
\begin{equation}
{\cal H}=\sum_{\langle i,j\rangle}  h_{ij} \label{fullHam}
\end{equation}
with  nearest neighbor sites $\langle i,j\rangle$.

Table \ref{table1} shows the most general basis vectors of the
space of local interactions for spin $S=2$ that are consistent
with the symmetries in terms of $S^z_i$-eigenstates
\begin{equation}
S_i^z\kets{m}_i =m_i\kets{m}_i \quad \text{ with } m \in
\{0,\pm1,\pm2\}. \label{eq-spin2}
\end{equation}
of the local spin-2 Hilbert space. It contains seven parameters
$a_1,\ldots,a_7$ that correspond to rotations of the basis vectors
in the respective subspaces that do not violate the symmetry
requirements.
\begin{table}
\caption{The symmetry-consistent spin-2 basis states in the space
of local interactions. Here $\kets{m,\tilde{m}}_\pm$ denotes the
(anti-) symmetrization $\kets{m,\tilde{m}}_\pm =\kets{m,\tilde{m}}
\pm \kets{\tilde{m},m}$, $\mu=m+\tilde{m}$ and $p=\pm1$ for
symmetric and antisymmetric states. \label{table1}}
\begin{scriptsize}
\begin{tabular}{rrll}
\hline$\mu$&p&name&state\\\hline
4&                  1&$\kets{v_4}                  $&$\kets{2,2}$\\
-4&       1&$\kets{v_{-4} }     $&$\kets{-2,-2}$\\
3&    1&$\kets{v_3^+}    $&$\kets{2,1}_+   :=\kets{2,1}+\kets{1,2}$\\
3&-1&$\kets{v_3^-   }       $&$\kets{2,1}_-:=\kets{2,1}-\kets{1,2}$\\
-3&      1&$\kets{v_{-3}^+}      $&$\kets{-2,-1}_+$\\
-3&-1&$\kets{v_{-3}^-}      $&$\kets{-2,-1}_-     $\\
2&     1&$\kets{v_{21}^+ }    $&$a_1     \kets{1,1}+     \kets{2,0}_+$\\
2&       1&$\kets{v_{22}^+}      $&$2\kets{1,1}-a_1\kets{2,0}_+$      \\
2&-1&$\kets{v_2^-}                 $&$\kets{2,0}_-$\\
-2&1&$\kets{v_{-21}^+}$&$a_1\kets{-1,-1}+\kets{-2,0}_+$\\
-2&1&$\kets{v_{-22}^+}$&$2\kets{-1,-1}-a_1\kets{-2,0}_+$\\
-2&-1&$\kets{v_{-2}^-}$&$\kets{-2,0}_-$\\
1&1&$\kets{v_{11}^+}$&$    a_2    \kets{1,0}_++\kets{2,-1}_+$\\
1&1&$\kets{v_{12}^+}$&$   \kets{1,0}_+   -a_2  \kets{2,-1}_+$\\
1&-1&$\kets{v_{11}^-}$&$a_3       \kets{1,0}_-+\kets{2,-1}_-$\\
1&-1&$\kets{v_{12}^-}$&$\kets{1,0}_-    -a_3   \kets{2,-1}_-$\\
-1&1&$\kets{v_{-11}^+}$&$a_2\kets{-1,0}_+ +\kets{-2,1}_+$\\
-1&1& $\kets{v_{-12}^+}$&$\kets{-1,0}_+ -a_2\kets{-2,1}_+$\\
-1&-1&$\kets{v_{-11}^-}$&$a_3 \kets{-1,0}_-+\kets{-2,1}_-$\\
-1&-1&$\kets{v_{-12}^-}$&$\kets{-1,0}_- - a_3\kets{-2,1}_-$\\
0&1&$\kets{v_{01}^+}$&$a_5 \kets{0,0}+a_4 \kets{1,-1}_+ +\kets{2,-2}_+$\\
0&1&$\kets{v_{02}^+}$&$ 2\frac{1+a_4 a_6}{a_5} \kets{0,0} -a_6\kets{1,-1}_+ -\kets{2,-2}_+$\\
0&1&$\kets{v_{03}^+}$&$\frac{-2a_5 (a_4 -a_6)}{2a_4 +(2a_4^2+a_5^2 )a_6} \kets{0,0}+ \frac{2+a_5^2+2a_4a_6}{2a_4+(2a_4^2+a_5^2)a_6} \kets{1,-1}_+-\kets{2,-2}_+$\\
0&-1&$\kets{v_{01}^-}$&$a_7 \kets{1,-1}_-+\kets{2,-2}_- $\\
0&-1&$\kets{v_{02}^-}$&$\kets{1,-1}_--a_7\kets{2,-2}_-$\\
\hline
\end{tabular}
\end{scriptsize}
\end{table}

The general form of the local interaction $h_{ij}$ can then be
expressed through the projectors on the symmetry-consistent basis
vectors:
\begin{eqnarray}
h_{ij}&=&\lambda_4(\kets{v_4}\bras{v_4}+\kets{v_{-4}}\bras{v_{-4}})+
\nonumber\\
&&\lambda_3^+(\kets{v_3^+}\bras{v_3^+}+\kets{v_{-3}^+}\bras{v_{-3}^+})+
\nonumber\\
&&\lambda_3^-(\kets{v_3^-}\bras{v_3^-}+\kets{v_{-3}^-}\bras{v_{-3}^-})+
\nonumber\\
&&\lambda_{21}^+(\kets{v_{21}^+}\bras{v_{21}^+}+\kets{v_{-21}^+}
\bras{v_{-21}^+})+\nonumber\\
&&\lambda_{22}^+(\kets{v_{22}^+}\bras{v_{22}^+}+\kets{v_{-22}^+}
\bras{v_{-22}^+})+\nonumber\\
&&\lambda_2^-(\kets{v_{2}^-}\bras{v_{2}^-}+\kets{v_{-2}^-}\bras{v_{-2}^-})
+\nonumber\\
&&\lambda_{11}^+(\kets{v_{11}^+}\bras{v_{11}^+}+\kets{v_{-11}^+}
\bras{v_{-11}^+})+\nonumber\\
&&\lambda_{12}^+(\kets{v_{12}^+}\bras{v_{12}^+}+\kets{v_{-12}^+}
\bras{v_{-12}^+})+\nonumber\\
&&\lambda_{11}^-(\kets{v_{11}^-}\bras{v_{11}^-}+\kets{v_{-11}^-}
\bras{v_{-11}^-})+\nonumber\\
&&\lambda_{12}^-(\kets{v_{12}^-}\bras{v_{12}^-}+\kets{v_{-12}^-}
\bras{v_{-12}^-})+\nonumber\\
&&\lambda_{01}^+\kets{v_{01}^+}\bras{v_{01}^+}+\lambda_{02}^+\kets{v_{02}^+}
\bras{v_{02}^+}+\nonumber\\
&&\lambda_{03}^+\kets{v_{03}^+}\bras{v_{03}^+}+\nonumber\\
&&\lambda_{01}^-\kets{v_{01}^-}\bras{v_{01}^-}+\lambda_{02}^-\kets{v_{02}^-}
\bras{v_{02}^-}. \label{eq-localh}
\end{eqnarray}
In the following we shall assume for convenience that the ground
state energy of the local interaction $h_{ij}$ is zero, i.e.\ at
least one $\lambda$ is equal to zero. The most general Hamiltonian
consistent with the symmetry requirements then has 21 free
parameters (seven $a_j$ and 15 $\lambda$-parameter, one of which
is used to fix the local ground state energy), including one
trivial scale factor.

Now the full Hamiltonian (\ref{fullHam}) is positive-semidefinite
since it is the sum of positive-semidefinite operators $h_{ij}$.
Therefore its ground state energy $E_0$ satisfies $E_0\geq 0$. By
definition an {\em optimum ground state} $\kets{\psi_0}$ has
$E_0=0$.\footnote{For the case where the local ground state energy
is
  $e_0\neq 0$, the global ground state energy scales with the number
  $N$ of bonds as $E_0=Ne_0$.}
It can be characterized by the condition
\begin{equation}
h_{ij}\kets{\psi_0}=0 \text{\ \  for  all\ }\langle i, j\rangle
\quad  \Longleftrightarrow \quad  {\cal H}\kets{\psi_0}=0,
\end{equation}
which means that  $\kets{\psi_0}$ is simultaneously ground state
of all local interactions $h_{ij}$. Therefore finite-size
corrections for the ground state energy vanish exactly for all
system sizes.

The simplest examples for OGS are tensor product states, e.g.\ the
fully-polarized ferromagnet
\begin{equation}
\kets{\psi_{\text{ferro}}}=\prod_{i}\kets{2}_i.
\end{equation}
The corresponding local ground state is $\kets{v_4}=\kets{2,2}$
(see Table 1). It will become a global ground state if one chooses
all interaction parameters $\lambda_{\mu n}^p$ in
(\ref{eq-localh}) as positive except for $\lambda_4=0$. The
parameters $a_1,\ldots,a_7$ are arbitrary.

\begin{figure}[ht]
\begin{center}
\epsfig{file=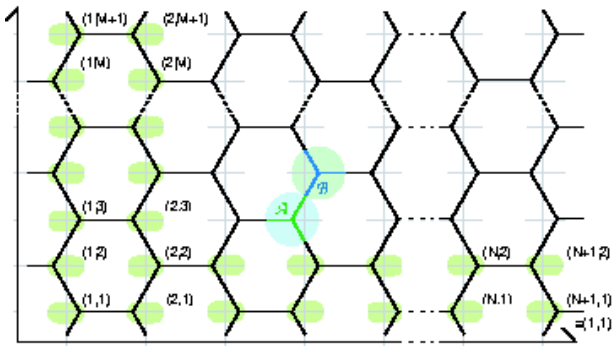,width=9cm} \caption{Numbering scheme for the
sites of the hexagonal lattice with sublattices $\mathfrak{A}$ and
$\mathfrak{B}$. \label{fig:gitter}}
\end{center}
\end{figure}

Here we are interested in more complex (antiferromagnetic) ground
states that are not given by simple tensor products. We shall now
consider a hexagonal lattice with periodic boundary conditions.
The specific implementation is depicted in Fig.~\ref{fig:gitter}
which also shows the numbering scheme for the sites. The hexagonal
lattice is bipartite with sublattices $\mathfrak{A,B}$.

\subsubsection{Vertex State Model approach}

On the hexagonal lattice there are two types of sites, belonging
to the two different sublattices $\mathfrak{A,B}$ (see
Fig.~\ref{fig:gitter}). Correspondingly two types of vertices can
be distinguished to which we assign quantum states
\begin{eqnarray}
\parbox[c][1.2cm][c]{1.2cm}{\epsfig{file=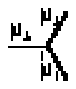,width=1cm}}
\hat{=}\
\alpha^1_{\mu_1,\mu_2,\mu_3}\kets{S^z_i(\mu_1,\mu_2,\mu_3)},
\ \  \text{for all }i \in \mathfrak{A},\label{vertexdef1}\\
\parbox[c][1.2cm][c]{1.2cm}{\epsfig{file=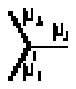,width=1cm}}
\hat{=}\
\alpha^2_{\mu_1,\mu_2,\mu_3}\kets{S^z_i(\mu_1,\mu_2,\mu_3)}, \ \
\text{for all\ } i \in \mathfrak{B},\label{vertexdef2}
\end{eqnarray}
i.e.\ the state vector $\kets{S^z_i}$ is determined by the values
$\mu_\alpha$ $(\alpha=1,2,3)$ of the bond variables belonging to
vertex $i$. In the following, $\mu_\alpha$ is a 2-state variable
that can be depicted graphically as an arrow pointing in or out of
the lattice site.

This general scheme of identifying vertices and quantum states is
rather natural. More specifically in the following we use the
identification
\begin{equation}
S_i^z=n_i + \frac{1}{2}\quad \text{or}\quad S_i^z=n_i -
\frac{1}{2} \label{eq-Szdef}
\end{equation}
with
\begin{equation}
 n_i:=\frac{1}{2}\left(n_i^{(+)} - n_i^{(-)}\right),
\end{equation}
where  $n_i^{(+)}$ and $n_i^{(-)}$ are the number of arrows
pointing out of and into vertex $i$, respectively. Therefore all
quantum states generated this way are $S^z$-eigenstates of the
local spin-2 Hilbert space (see eq.~(\ref{eq-spin2})).

There is still some freedom in the choice of the sign in the
additive constant eq.~(\ref{eq-Szdef}). Since here we are
interested in ground states that are not macroscopically
degenerate it will be fixed for each sublattice (see
Sec.~\ref{sec:model}). For the same reason we will not consider
bond variables $\mu_\alpha$ that can take more than two
states\footnote{1-state bond variables $\mu_\alpha$ lead to simple
tensor products only.} which is possible in principle. The
prefactors $\alpha^i_{\mu_1,\mu_2,\mu_3}$  are not determined by
this method and can be chosen in a suitable way restricted only by
the required symmetries.

In the next step we define quantum states of the {\em full}
lattice from the single-site states introduced above. As in the
partition function of a classical vertex model the bond variables
are summed out. For two sites this implies
\begin{equation}
\begin{tabular}{l}
\parbox[c][1.2cm][c]{.9cm}{\epsfig{file=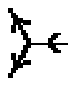,height=1cm}}
$\otimes$\parbox[c][1.2cm][c]{.9cm}{
\epsfig{file=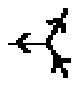,height=1cm}}$+$
\parbox[c][1.2cm][c]{.9cm}{\epsfig{file=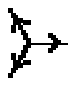,height=1cm}}
$\otimes$\parbox[c][1.2cm][c]{.9cm}{
\epsfig{file=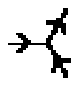,height=1cm}}=\\
\parbox[c][1.3cm][c]{1.5cm}{\epsfig{file=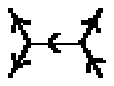,height=1cm}}$+$
\parbox[c][1.3cm][c]{1.5cm}{\epsfig{file=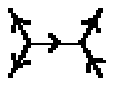,height=1cm}}=
\parbox[c][1.3cm][c]{1.4cm}{\epsfig{file=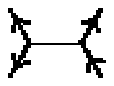,height=1cm}}
\end{tabular}
\label{eq-2spinstates}
\end{equation}
or, explicitly,
\begin{equation}
\alpha^1_1\alpha^2_2\kets{1,1}+\alpha^1_2\alpha^2_2\kets{2,1}=
\kets{\psi^{1,2}}.
\end{equation}
Here the only difference is replacing the product of Boltzmann
weights by a tensor product in spin space. This procedure is
generalized to any number of sites in a straightforward way to
construct a quantum state on the hexagonal lattice consisting of
local spin-2 variables:
\begin{equation}
\kets{\psi_0}=\sum_{\{\mu_i\}}\left(\prod_{i\in
\mathfrak{A}}^\otimes
\parbox[c][1.2cm][c]{1.1cm}{\epsfig{file=2.eps,width=1cm}}\right)\otimes\left(\prod_{i\in
\mathfrak{B}}^\otimes
\parbox[c][1.2cm][c]{1.1cm}{\epsfig{file=3.eps,width=1cm}}\right),
\label{eq-ground}
\end{equation}
where the black vertices have to be taken for $i\in \mathfrak{A}$
and the grey ones for $i\in \mathfrak{B}$.

So far this procedure is independent of the Hamiltonian under
consideration, except for the choice of the local Hilbert space.
In the final step we now have to specify the free parameters in
the local interaction (\ref{eq-localh}) and the wavefunction
(\ref{vertexdef1}), (\ref{vertexdef2}) in such a way that the
quantum state (\ref{eq-ground}) becomes (i) an eigenstate and (ii)
a ground state of the Hamiltonian (\ref{fullHam}). This can be
achieved by imposing the optimum ground state condition
\begin{equation}
h_{ij}\left[\parbox[c][1.1cm][c]{1.5cm}{\epsfig{file=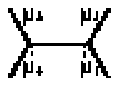,height=1.2cm}}\right]=h_{ij}\left[\parbox[c][1.1cm][c]{1.4cm}{\epsfig{file=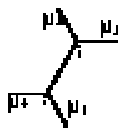,height=1.7cm}}\right]=h_{ij}\left[\parbox[c][1.2cm][c]{1.3cm}{\epsfig{file=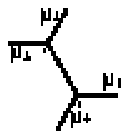,height=1.7cm}}\right]=0,
\end{equation}
for any pair of nearest neighbors $\langle i,j\rangle$ and all
values of $\mu_\alpha$ for $\alpha=1,2,3,4$. Since $h_{ij}$ is
positive-semidefinite this means that the corresponding 2-spin
states (\ref{eq-2spinstates}) are ground states of the local
interaction. Therefore, by construction, the global ground state
(\ref{eq-ground}) contains only local ground states and thus the
optimum ground state condition ${\cal H}\kets{\psi_0}=0$ is
fulfilled.


\section{Ground states}
\label{sec:model}

In the following we will discuss explicit results for ground state
wave functions. By specifying the sign of additive constant in
(\ref{eq-Szdef}) we find two different types of states.

\subsection{The Weak Antiferromagnet}

First consider the case where the additive constant in
(\ref{eq-Szdef}) is fixed to $+\frac{1}{2}$ for sublattice
$\mathfrak{A}$ and to $-\frac{1}{2}$ for sublattice
$\mathfrak{B}$, e.g.\
\parbox[c][1.2cm][c]{.9cm}{\epsfig{file=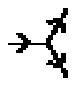,height=1cm}}
$\hat{=}\alpha_1\kets{1}$ and
\parbox[c][1.2cm][c]{.9cm}{\epsfig{file=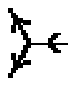,height=1cm}}
$\hat{=}  \alpha_2 \kets{0}$. The freedom in choosing the sign in
(\ref{eq-Szdef}) can be interpreted as a fourth 2-state variable
$\mu_4$ attached to the site (see
(\ref{vertexdef1}),(\ref{vertexdef2})) which is not summed over in
the construction of the global state (\ref{eq-ground}). It
corresponds to a local spin-1/2 degree of freedom with
magnetization $+\frac{1}{2}$ on sublattice $\mathfrak{A}$ and
$-\frac{1}{2}$ on sublattice $\mathfrak{B}$. In this sense we can
view the construction described above as starting from a
N\'eel-type reference state with sublattice magnetization
$\frac{1}{2}$ which is modified by taking into account the (local)
vertex configurations. Therefore we have the following
identification of vertices and local quantum states:
\begin{equation}
\begin{tabular}{|ccr|ccr|ccr|ccr|}\hline
 \epsfig{file=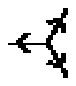} & $\hat{=}$&$              a \kets{2} $&
 \epsfig{file=2r.eps} & $\hat{=}$&$              b \kets{1} $&
 \epsfig{file=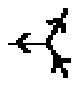} & $\hat{=}$&$              b \kets{1} $&
 \epsfig{file=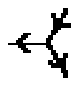} & $\hat{=}$&$              b \kets{1} $\\\hline
 \epsfig{file=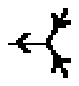} & $\hat{=}$&$              c \kets{0} $&
 \epsfig{file=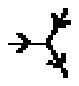} & $\hat{=}$&$              c \kets{0} $&
 \epsfig{file=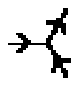} & $\hat{=}$&$              c \kets{0} $&
 \epsfig{file=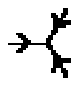} & $\hat{=}$&\hspace{-0.1cm}$              b \kets{\text{-}1} $\\\hline
 \epsfig{file=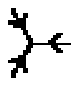} & $\hat{=}$&$              a \kets{\text{-}2} $&
 \epsfig{file=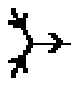} & $\hat{=}$&\hspace{-0.1cm}$      \sigma  b \kets{\text{-}1}$&
 \epsfig{file=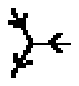} & $\hat{=}$&\hspace{-0.1cm}$      \sigma  b \kets{\text{-}1} $&
 \epsfig{file=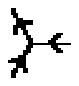} & $\hat{=}$&\hspace{-0.1cm}$      \sigma  b \kets{\text{-}1}$\\\hline
 \epsfig{file=2l.eps} & $\hat{=}$&$              c \kets{0} $&
 \epsfig{file=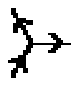} & $\hat{=}$&$              c \kets{0}$&
 \epsfig{file=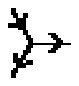} & $\hat{=}$&$              c \kets{0}$&
 \epsfig{file=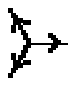} & $\hat{=}$&$      \sigma  b \kets{1}$\\ \hline
\end{tabular}
\nonumber
\end{equation}
They contain three real parameters $a,b,c \in \mathbb{R}$ and one
discrete parameter $\sigma  =\pm 1$. One of the real parameters
can always be absorbed into the spectral parameters $\lambda$ in
(\ref{eq-localh}). This choice of parameters is the most general
one allowed by symmetry requirements.

Summing out an inner bond variable leads to nine 2-spin states
which become local ground states in the following:
\begin{align}
  ac \kets{2,0} +  \sigma b^2   \kets{1,1}      \nonumber    \\
  ac \kets{0,-2}+  \sigma b^2   \kets{-1,-1}    \nonumber    \\
  \sigma ab \kets{2,-1}+  bc    \kets{1,0}      \nonumber    \\
  ab \kets{1,-2}+  bc \sigma    \kets{0,-1}     \nonumber    \\
  bc\kets{1,0}_\sigma                \label{eq-waf}               \\
  bc\kets{-1,0}_\sigma                          \nonumber    \\
  a^2\kets{-2,2}+ \sigma b^2    \kets{-1,1}     \nonumber    \\
  \sigma b^2\kets{-1,1}+ c^2    \kets{0,0}      \nonumber    \\
  \sigma b^2\kets{1,-1}+ c^2    \kets{0,0}      \nonumber
\end{align}
As the system should be invariant under time reversal the
spin-flipped reference state, which corresponds to an exchange of
the sublattices, should also lead to a ground state. This second
ground state can be constructed by the following identification of
vertices and local states:
\begin{equation}
\begin{tabular}{|ccr|ccr|ccr|ccr|}\hline
    \epsfig{file=1r.eps} & $\hat{=}$&$              b \kets{1} $&
    \epsfig{file=2r.eps} & $\hat{=}$&$              c \kets{0} $&
    \epsfig{file=3r.eps} & $\hat{=}$&$              c \kets{0} $&
    \epsfig{file=4r.eps} & $\hat{=}$&$              c \kets{0} $\\\hline
    \epsfig{file=5r.eps} & $\hat{=}$&\hspace{-0.1cm}$ b \kets{\text{-}1} $&
    \epsfig{file=6r.eps} & $\hat{=}$&\hspace{-0.1cm}$ b \kets{\text{-}1} $&
    \epsfig{file=7r.eps} & $\hat{=}$&\hspace{-0.1cm}$ b \kets{\text{-}1} $&
    \epsfig{file=8r.eps} & $\hat{=}$&\hspace{-0.1cm}$ a \kets{\text{-}2} $\\\hline
    \epsfig{file=8l.eps} & $\hat{=}$&$      \sigma  b \kets{\text{-}1} $&
    \epsfig{file=5l.eps} & $\hat{=}$&\hspace{-0.1cm}$      c \kets{0}$&
    \epsfig{file=6l.eps} & $\hat{=}$&\hspace{-0.1cm}$      c \kets{0} $&
    \epsfig{file=7l.eps} & $\hat{=}$&\hspace{-0.1cm}$      c \kets{0}$\\\hline
    \epsfig{file=2l.eps} & $\hat{=}$&$      \sigma  b \kets{1} $&
    \epsfig{file=3l.eps} & $\hat{=}$&$      \sigma  b \kets{1}$&
    \epsfig{file=4l.eps} & $\hat{=}$&$      \sigma  b \kets{1}$&
    \epsfig{file=1l.eps} & $\hat{=}$&$              a \kets{2}$\\ \hline
\end{tabular}
\nonumber
\end{equation}
As a result the following five additional local ground states are
obtained:
\begin{align}
           ac \kets{0,2} +  \sigma b^2       \kets{1,1}   \nonumber    \\
           ac \kets{-2,0}+  \sigma b^2       \kets{-1,-1} \nonumber    \\
           a^2\kets{2,-2}+ \sigma b^2    \kets{1,-1}                       \\
           \sigma ab \kets{-1,2}+  bc        \kets{0,1}   \nonumber    \\
           ab \kets{-2,1}+  bc \sigma        \kets{-1,0}. \nonumber
\end{align}
The total number of local ground states is therefore 14. The
following conditions must be satisfied:
\begin{equation}
\lambda_{2}^-=\lambda_{22}^+=\lambda_{11}^\sigma=\lambda_{12}^+
=\lambda_{12}^-=\lambda_{02}^+=\lambda_{03}^+=\lambda_{01}^-=\lambda_{02}^-=0
\label{eq-gleich}
\end{equation}
and
\begin{equation}
\lambda_{4} ,\lambda_{3}^+ ,\lambda_{3}^- ,\lambda_{21}^+
,\lambda_{11}^{-\sigma} ,\lambda_{01}^+ > 0, \label{eq-ungleich}
\end{equation}
where $\sigma$ is the discrete parameter appearing in the
identification of vertices with quantum states. The equations
(\ref{eq-gleich}) guarantee that the states are eigenstates
whereas the inequalities (\ref{eq-ungleich}) are needed to make
them ground states.

The global quantum state (\ref{eq-ground}) is the twofold
degenerate ground state for the Hamiltonian (\ref{fullHam}),
(\ref{eq-localh}) satisfying above conditions (\ref{eq-gleich}),
(\ref{eq-ungleich}). The degeneracy is exact for all lattice
sizes. We do not expect the existence of a phase transition within
the parameter region where the optimum ground state is realized.

The global ground state is effectively controlled by two
continuous parameters $\frac{a}{c}, \frac{b}{c}$ and one discrete
one $\sigma\pm 1$. The global Hamiltonian can be tuned by a
parameter manifold of eight continuous (namely the six parameters
in (\ref{eq-ungleich}) and $a/c$, $b/c$) and one discrete
parameter $\sigma$ with still a free choice of the scale.
 By construction, the ground state is antiferromagnetically ordered. From the reference
state it inherits a N\'eel-type order. The sublattice
magnetizations $\mathfrak{m}_{\mathfrak{A}}=\langle S_i^z\rangle$
for $i\in\mathfrak{A}$ and $\mathfrak{m}_{\mathfrak{B}}=\langle
S_i^z\rangle$ for $i\in\mathfrak{B}$ depend on the parameters
$\frac{a}{c}, \frac{b}{c}$ and $\sigma=\pm 1$ and satisfy
\begin{equation}
\mathfrak{m}:=\mathfrak{m}_{\mathfrak{A}}=-\mathfrak{m}_{\mathfrak{B}}
\quad \text{with}\quad 0\leq \mathfrak{m}\leq 2.
\end{equation}
Therefore in general the sublattices are not fully polarized
($\mathfrak{m} <2$) and we call the state a {\em weak
antiferromagnet}. For $|a|\to\infty$ one can see from
(\ref{eq-waf}) that the local and global ground states are the
N\'eel-states with $\mathfrak{m}=2$. On the other hand, for
$|c|\to\infty$ the ground states becomes $|0\cdots0\rangle$ and
has $\mathfrak{m}=0$.

\subsection{The Weak Ferromagnet}\label{sec:WF}

It is also possible to start from a different reference state by
chosing the same sign for both sublattices in (\ref{eq-Szdef}).
This state is ferromagnetically ordered with magnetization
$\frac{1}{2}$. Using the same conventions as above, we obtain the
following identification of vertices with local quantum states:
\begin{equation}
\begin{tabular}{|ccr|ccr|ccr|ccr|}\hline
    \epsfig{file=1r.eps} & $\hat{=}$&$              a \kets{2} $&
    \epsfig{file=2r.eps} & $\hat{=}$&$              b \kets{1} $&
    \epsfig{file=3r.eps} & $\hat{=}$&$              b \kets{1} $&
    \epsfig{file=4r.eps} & $\hat{=}$&$              b \kets{1} $\\\hline
    \epsfig{file=5r.eps} & $\hat{=}$&$              c \kets{0} $&
    \epsfig{file=6r.eps} & $\hat{=}$&$              c \kets{0} $&
    \epsfig{file=7r.eps} & $\hat{=}$&$              c \kets{0} $&
    \epsfig{file=8r.eps} & $\hat{=}$&\hspace{-0.1cm}$      b \kets{\text{-}1} $\\\hline
    \epsfig{file=8l.eps} & $\hat{=}$&$      \sigma  b \kets{\text{-}1} $&
    \epsfig{file=5l.eps} & $\hat{=}$&\hspace{-0.1cm}$      c \kets{0}$&
    \epsfig{file=6l.eps} & $\hat{=}$&\hspace{-0.1cm}$      c \kets{0} $&
    \epsfig{file=7l.eps} & $\hat{=}$&\hspace{-0.1cm}$      c \kets{0}$\\\hline
    \epsfig{file=2l.eps} & $\hat{=}$&$      \sigma  b \kets{1} $&
    \epsfig{file=3l.eps} & $\hat{=}$&$      \sigma  b \kets{1}$&
    \epsfig{file=4l.eps} & $\hat{=}$&$      \sigma  b \kets{1}$&
    \epsfig{file=1l.eps} & $\hat{=}$&$              a \kets{2}$.\\ \hline
\end{tabular}
\nonumber
\end{equation}
These vertices lead to nine local 2-spin states:
\begin{align}
 ab \kets{2,1}_\sigma                   \nonumber  \\
 ac \kets{2,0}+  \sigma b^2 \kets{1,1}  \nonumber  \\
 ac \kets{0,2}+  \sigma b^2 \kets{1,1}  \nonumber  \\
 \sigma ab \kets{2,-1}+  bc \kets{1,0}  \nonumber  \\
 ab \kets{-1,2}+  \sigma bc \kets{0,1}  \label{eq-wf1} \\
 bc\kets{1,0}_\sigma                    \nonumber  \\
 c^2\kets{0,0}+ \sigma b^2  \kets{-1,1} \nonumber  \\
 c^2\kets{0,0}+ \sigma b^2  \kets{1,-1} \nonumber  \\
 bc\kets{-1,0}_\sigma       ,           \nonumber
\end{align}
where $a,b,c$ are real and $\sigma=\pm 1$ is a discrete parameter.

Again the global ground state is obtained by summing over all
inner bond variables. The spin-flipped reference state with
magnetization $-\frac{1}{2}$ will also lead to a ground state if
the following additional vertices are used:
\begin{equation*}
\begin{tabular}{|ccr|ccr|ccr|ccr|}\hline
    \epsfig{file=1r.eps} & $\hat{=}$&$              b \kets{1} $&
    \epsfig{file=2r.eps} & $\hat{=}$&$              c \kets{0} $&
    \epsfig{file=3r.eps} & $\hat{=}$&$              c \kets{0} $&
    \epsfig{file=4r.eps} & $\hat{=}$&$              c \kets{0} $\\\hline
    \epsfig{file=5r.eps} & $\hat{=}$&$              b \kets{\text{-}1} $&
    \epsfig{file=6r.eps} & $\hat{=}$&$              b \kets{\text{-}1} $&
    \epsfig{file=7r.eps} & $\hat{=}$&$              b \kets{\text{-}1} $&
    \epsfig{file=8r.eps} & $\hat{=}$&\hspace{-0.1cm}$a \kets{\text{-}2}$\\\hline
    \epsfig{file=8l.eps} & $\hat{=}$&$                           a \kets{\text{-}2}$&
    \epsfig{file=5l.eps} & $\hat{=}$&\hspace{-0.1cm}$ \sigma     b \kets{\text{-}1} $&
    \epsfig{file=6l.eps} & $\hat{=}$&\hspace{-0.1cm}$ \sigma     b \kets{\text{-}1} $&
    \epsfig{file=7l.eps} & $\hat{=}$&\hspace{-0.1cm}$ \sigma     b \kets{\text{-}1}$\\\hline
    \epsfig{file=2l.eps} & $\hat{=}$&$              c \kets{0} $&
    \epsfig{file=3l.eps} & $\hat{=}$&$              c \kets{0}$&
    \epsfig{file=4l.eps} & $\hat{=}$&$              c \kets{0}$&
    \epsfig{file=1l.eps} & $\hat{=}$&$ \sigma       b \kets{1}$.\\ \hline
\end{tabular}
\end{equation*}
These vertices generate five additional 2-spin states:
\begin{align}
           ab \kets{-2,-1}_\sigma                     \nonumber\\
           ac \kets{-2,0}+  \sigma b^2  \kets{-1,-1}  \nonumber       \\
           ac \kets{0,-2}+  \sigma b^2  \kets{-1,-1} \label{eq-wf2}    \\
    \sigma ab \kets{-2,1}+  bc          \kets{-1,0}   \nonumber        \\
           ab \kets{1,-2}+  \sigma bc   \kets{0,-1}   \nonumber.
\end{align}
The 14 states (\ref{eq-wf1}) and (\ref{eq-wf2}) are local ground
states for
\begin{equation}
\lambda_{3}^\sigma=\lambda_{21}^+=\lambda_{2}^-=\lambda_{11}^-
=\lambda_{12}^\sigma=\lambda_{03}^+=\lambda_{02}^-=\lambda_{01}^-=0
\label{eq-equalwf}
\end{equation}
and
\begin{equation}
\lambda_{4} ,\lambda_{3}^{-\sigma}
,\lambda_{22}^{+},\lambda_{12}^{-\sigma} ,\lambda_{01}^{+}
,\lambda_{02}^+,\lambda_{01}^-> 0. \label{eq-inequalwf}
\end{equation}
Again, the corresponding global states (\ref{eq-ground}) are
controlled by two continuous parameters $\frac{a}{c}, \frac{b}{c}$
and one discrete one $\sigma\pm1$. The freedom of chosing the
reference state such that its magnetization is $+\frac{1}{2}$ or
$-\frac{1}{2}$ implies that a priori the ground state is exactly
twofold degenerate.

The global state generically has sublattice magnetizations $-1\leq
\mathfrak{m}_\mathfrak{B}\leq \frac{1}{2}\leq
\mathfrak{m}_\mathfrak{A} \leq 2$ or $1\geq
\mathfrak{m}_\mathfrak{B}\geq \frac{1}{2}\geq
\mathfrak{m}_\mathfrak{A} \geq -22$. It is convenient to introduce
the order parameter $\psi =\mathfrak{m} -\frac{1}{2}$ which is the
magnetization relative to the reference state (with additive
constant $+1/2$). Then we have
\begin{equation}
\psi = \psi_\mathfrak{A}=-\psi_\mathfrak{B}\quad \text{with}\quad
0 \leq \psi \leq \frac{3}{2}.
\end{equation}
Note that the magnetization $\mathfrak{m}_\mathfrak{B}$ on
sublattice $\mathfrak{B}$ can indeed become negative so that the
global ground state exhibits an antiferromagnetic ordering.

Most interestingly the structure of the ground state suggests the
occurrence of a quantum phase transition in the solvable subspace
(\ref{eq-equalwf}), (\ref{eq-inequalwf}) of the weakly
ferromagnetic ground state. If the product $ab$ of the parameters
controlling the local ground states becomes large compared to the
other coefficients in (\ref{eq-wf1}), the dominating 2-spin states
contributing to the global state are $ab\kets{2,1}_\sigma$,
$\sigma ab \kets{2,-1}$ and $ab \kets{-1,2}$ (and its spin-flipped
counterparts). However, the only global states compatible with
periodic boundary conditions that can be constructed from these
local states are product states having $\kets{2}$ on sublattice
$\mathfrak{A}$ and $\kets{-1}$ on sublattice $\mathfrak{B}$, or
vice versa. This implies a fourfold degenerate ground state with
sublattice symmetry in the limit $ab\to\infty$. In this limit the
order parameter becomes $\psi=\frac{3}{2}$.  On the other hand,
for small $ab$ the ground state is only twofold degenerate. Thus
we can expect a quantum phase transition at a critical value
$(ab)_c$ from a fourfold degenerate state to a twofold degenerate
state with vanishing order parameter $\psi =0$, i.e.\ with
non-broken sublattice symmetry. The determination of correlation
functions requires studying a two-dimensional classical vertex
model (see Sec.~\ref{sec:concl}). Preliminary
results\cite{inprep04} indeed show strong evidence for a phase
transition line. Numerical results for a representative point
indicate a continuous transition.


\section{Conclusions} \label{sec:concl}

In this paper we have solved a large class of spin-2 quantum spin
models on the hexagonal lattice. Using the optimum ground state
concept realized through the quantum vertex model approach the
explicit form of the ground state wave function could be
determined. Focussing on weakly degenerate ground states two
different types of states existing in different subspaces have
been found. In contrast to the generic VBS states
\cite{AKLT87,AKLT88}, coordination number $z$ and spin $S$ are not
related by $S=z/2$. Some properties of the states that we have
constructed can be understood by interpreting the spin $S=2$ as
symmetrization of a spin-3/2 and a spin-1/2. For the spin $S=3/2$
OGS have been constructed previously \cite{NKZ97}. In addition we
now have a spin-1/2 degree of freedom at each site that allows to
define different ``reference states''.

The weak antiferromagnet has a twofold degenerate ground state
characterized by a sublattice magnetization
$\mathfrak{m}_\mathfrak{A} =-\mathfrak{m}_\mathfrak{B}$ with
$0\leq \mathfrak{m}_\mathfrak{A}\leq 2$. It exists in a subspace
controlled by eight continuous and one discrete parameter. The
weak ferromagnet is conveniently described by the order parameter
$\psi =\mathfrak{m} -\frac{1}{2}$ such that
$\psi_\mathfrak{A}=-\psi_\mathfrak{B}$. We have argued that it
exhibits a quantum phase transition from a twofold degenerate
disordered phase with $\psi = 0$ to a fourfold degenerate phase
with broken sublattice symmetry and $\psi > 0$.

In future work we will investigate the ground state correlations
which will help to understand the properties of the quantum phase
transition in more detail. As described in \cite{NKZ97}
expectation values and correlations are determined by the
partition function of a {\em classical} vertex model on the same
lattice. This is a considerable simplification compared to the
generic case where e.g.\ Quantum Monte Carlo calculations require
the investigation of a classical model in dimension $d+1$.
However, in our case the classical vertex model has four states
per bond and is not exactly solvable. Therefore its properties
have to be studied numerically using {\em classical} Monte Carlo
simulations.

We want to emphasize that the construction of quantum states using
vertex models is different from the one used in so-called quantum
vertex models \cite{Ardonne04}. There the graphical representation
of the states in terms of closed loops is used to define the full
dynamics of a quantum model, i.e.\ the full Hamiltonian. In our
case the vertex model is only used to specify the ground state.
The identification of vertex and quantum states does not extend to
excitations. This offers a larger flexibility of the approach.


\begin{acknowledgments}
  This work has been performed within the research program SFB 608 of
  the \emph{Deutsche Forschungsgemeinschaft}.
\end{acknowledgments}

\bibliography{VPGS}

\end{document}